\documentclass[journal]{IEEEtran}

\usepackage[utf8]{inputenc}
\usepackage[dvipsnames]{xcolor}
\usepackage[caption=false]{subfig}
\usepackage{color}
\usepackage{graphicx}
\usepackage{balance}
\usepackage{amsmath}
\usepackage{amsfonts}
\usepackage{amssymb}
\usepackage{pifont}
\usepackage{times}
\usepackage{paralist}
\usepackage{textcomp}
\usepackage[hyphens]{url}
\usepackage{siunitx}
\usepackage{paralist}
\usepackage{amsmath,colortbl}
\usepackage{cite,comment}
\usepackage[nolist]{acronym}
\usepackage{soul}
\usepackage{array}
\usepackage{tabularx}
\usepackage{multirow}
\usepackage{pgfplots}
\usepackage{enumitem}
\usepackage{makecell}
\usepackage[flushleft]{threeparttable}
\usepackage{tikz}
\usetikzlibrary{arrows.meta,
                backgrounds,
                chains,
                fit,
                quotes}
\usetikzlibrary{positioning}

\pgfplotsset{compat=1.18}
\usepackage{stackengine}
\newcolumntype{P}[1]{>{\centering\arraybackslash}m{#1}}
\newcolumntype{Q}[1]{>{\arraybackslash}m{#1}}
\newcommand{\counting}[1]{#1} 

\begin{acronym}[ACRONYM]
\acro{3GPP}{the third generation partnership project}
\acro{5G}{fifth generation}
\acro{6G}{sixth generation}
\acro{AI}{artificial intelligence}
\acro{AoA}{angle-of-arrival}
\acro{AoD}{angle-of-departure}
\acro{BS}{base station}
\acro{D-MIMO}{distributed MIMO}
\acro{EM}{electromagnetic}
\acro{GDOP}{geometric dilution of precision}
\acro{GNSS}{global navigation satellite system}
\acro{IQ}{in-phase and quadrature}
\acro{DCC}{dynamic cooperation clustering}
\acro{JCS}{Joint Communication and Sensing}
\acro{JRC}{joint radar and communication}
\acro{JRC2LS}{joint radar communication, computation, localization, and sensing}
\acro{ICI}{inter-carrier interference}
\acro{IOO}{indoor open office}
\acro{IoT}{Internet of Things}
\acro{IRN}{infrastructure reference node}
\acro{KPI}{key performance indicator}
\acro{MRC}{maximum ratio combining}
\acro{LOS}{line-of-sight}
\acro{LAS}[L\&S]{localization and sensing}
\acro{MIMO}{multiple-input multiple-output}
\acro{mmWave}{millimeter-wave}
\acro{NLOS}{non-line-of-sight}
\acro{NR}{new radio}
\acro{OFDM}{orthogonal frequency-division multiplexing}
\acro{OTFS}{orthogonal time-frequency-space}
\acro{PRS}{positioning reference signal}
\acro{QoS}{Quality of Service}
\acro{RAN}{radio access network}
\acro{RAT}{radio access technology}
\acro{RedCap}{reduced capacity}
\acro{RIS}{reconfigurable intelligent surface}
\acro{RTK}{real-time kinematic}
\acro{RTT}{round-trip-time}
\acro{SLAM}{simultaneous localization and mapping}
\acro{SNR}{signal-to-noise ratio}
\acro{SOTA}{state of the art}
\acro{ToA}{time-of-arrival}
\acro{TDoA}{time-difference-of-arrival}
\acro{TDD}{time division duplex}
\acro{TR}{time-reversal}
\acro{TRP}{transmission and reception point}
\acro{TXRX}[TX/RX]{transmitter/receiver}
\acro{TX}{transmitter}
\acro{RX}{receiver}
\acro{UE}{user equipment}
\acro{multi-RTT}{multi-cell round-trip-time}
\acro{UL-TDOA}{uplink time-difference-of-arrival}
\acro{DL-TDOA}{downlink time-difference-of-arrival}
\acro{UMi}{3D-urban micro}
\acro{UMa}{3D-urban macro}
\acro{FR1}{frequency range 1}
\acro{FR2}{frequency range 2}
\acro{ISAC}{integrated sensing and communication}
\acro{ICIC}{inter-cell interference coordination}
\acro{CoMP}{coordinated multipoint}
\acro{IAB}{integrated access and backhaul}
\acro{RIS}{reconfigurable intelligent surfaces }
\acro{NCR}{network-controlled repeater}
\acro{C-RAN}{cloud radio access network}
\acro{CSI}{channel state information}
\acro{CPU}{central processing unit}
\acro{AP}{access point}
\acro{CLOSE}{communication, localization, and sensing}
\acro{UL}{uplink}
\acro{DL}{downlink}
\acro{FH}{fronthaul}
\acro{BH}{backhaul}
\acro{RU}{radio unit}
\acro{DU}{distributed unit}
\acro{CU}{central unit}
\acro{FR}{frequency range}
\acro{SE}{spectral efficiency}
\acro{CJT}{coherent joint transmission}
\acro{TRP}{transmission reception point}
\acro{RMSE}{root-mean-square error}
\acro{PEB}{position error bound}
\end{acronym}

\begin{document}

\title{Integrated Communication, Localization, and Sensing in 6G D-MIMO Networks}

\author{Hao Guo, Henk Wymeersch, Behrooz Makki, Hui Chen, Yibo Wu, Giuseppe Durisi, \\Musa Furkan Keskin, Mohammad H. Moghaddam, Charitha Madapatha, Han Yu, \\Peter Hammarberg, Hyowon Kim, and Tommy Svensson}

\twocolumn
\maketitle

\begin{abstract}
Future generations of mobile networks call for concurrent sensing and communication functionalities in the same hardware and/or spectrum. Compared to communication, sensing services often suffer from limited coverage, due to the high path loss of the reflected signal and the increased infrastructure requirements.  To provide a more uniform quality of service, distributed multiple input multiple output (D-MIMO) systems deploy a large number of distributed nodes and efficiently control them, making distributed integrated sensing and communications (ISAC) possible.  In this paper, we investigate  ISAC in D-MIMO through the lens of different design architectures and deployments, revealing both conflicts and synergies. In addition, simulation and demonstration results reveal both opportunities and challenges towards the implementation of ISAC in D-MIMO. 
\end{abstract}


\IEEEpeerreviewmaketitle

\section{Introduction}%
\label{section:intro}
With the success of \ac{MIMO}, it is expected that multi-antenna technologies will evolve in beyond-5G systems, either in a centralized or a distributed way. In the centralized case, the \acp{AP} or \acp{UE} will be equipped with an even larger number of antennas. In the distributed case, also referred to as \ac{D-MIMO}, multiple multi-antenna \acp{AP} with potentially different capabilities will cooperate to serve the \acp{UE}~\cite{cell-free1}. \ac{D-MIMO} is considered as a key element in the $6$th generation of mobile communications~\cite{haliloglu2023dmimosystem}. Unlike conventional \ac{MIMO}, where multiple antennas are concentrated at a single location, the distributed architecture of \ac{D-MIMO} facilitates a new level of spatial diversity and cooperative communication with a degree of freedom that enables, e.g., blockage avoidance and increased link margin despite per node output power limitations, leading to high reliability and availability as well as uniform service over the coverage area \cite{GMA+22}. Moreover, by spreading antennas across different locations, \ac{D-MIMO} offers inherent advantages in terms of, e.g., coverage, spectrum efficiency, and energy efficiency. The cooperative nature of \ac{D-MIMO} also enables the network to adapt more dynamically to channel variations, ensuring efficient use of resources and mitigating the impact of fading or other impairments, at the cost of more complicated backhaul/fronthaul structures \cite{cell-free2}. 

With these promising features, \ac{D-MIMO} can be an attractive solution for so-called \ac{ISAC}, where the same hardware and/or frequency bands are used to perform these functionalities in a distributed and cooperative way~\cite{MIMO-ISAC-arxiv}. Specifically, sensing includes mono-static and multi-static modes of operation, enabling services such as user localization, object detection and tracking, or gesture recognition, which build on these fundamental sensing modes. 

Traditionally, radar sensing and communication have operated in separate frequency bands using dedicated hardware. However, with 5G and beyond,  the wireless communication bands are merging with radar bands, such as \ac{mmWave} and the sub-THz bands foreseen for 6G. This merging has fueled the research on integrating communication, localization, and sensing functionalities within the same system, which can offer several benefits.  One major advantage of \ac{ISAC} is centralized resource allocation and interference management for all functionalities, leading to cost-efficient operations. With \ac{D-MIMO}'s distributed node characteristics, more flexibility is provided in resource allocation, including time, frequency, space, and energy, across sensing, localization, and communication signals.  Additionally, integrating sensing into the existing communication infrastructure reduces the cost significantly compared to deploying a separate sensing network. Moreover, leveraging multiple multi-antenna nodes increases the likelihood of \ac{LOS} links and provides the network with multiple perspectives on UEs/objects, thereby enhancing localization and sensing performance. The implementation of \ac{ISAC} also brings benefits to \ac{D-MIMO} networks. Specifically, \ac{LAS} enhance the network's radio environment comprehension, such as detecting blockages~\cite{GMA+22}. This knowledge simplifies backhaul/fronthaul designs and reduces coordination overheads, as only \acp{AP} with strong links to \acp{UE}/objects need to collaborate.

\begin{figure*}
    \centering
\includegraphics[width=1.99\columnwidth]{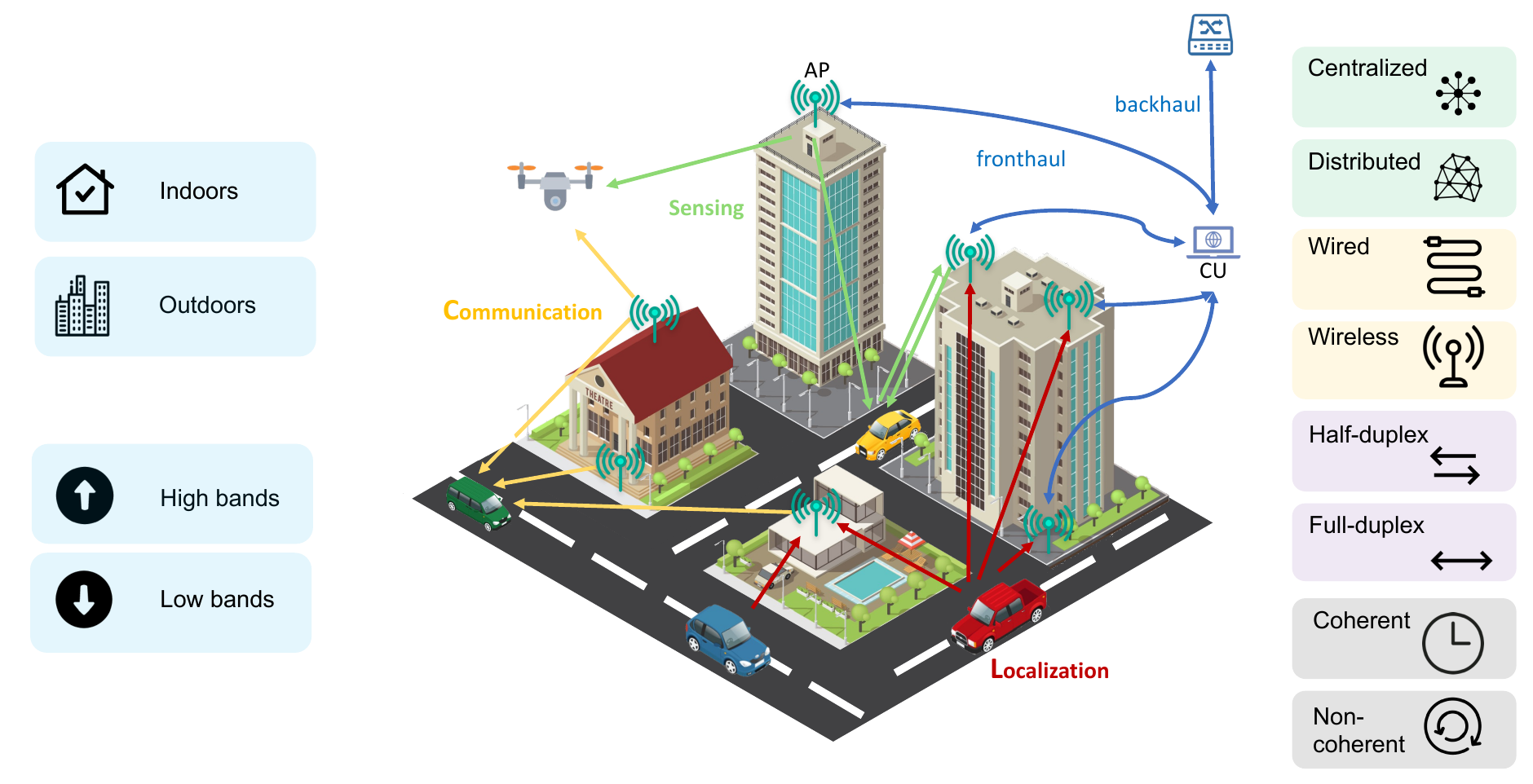}
\vspace{-0.3cm}
    \caption{Illustration of the ISAC functionalities in D-MIMO systems with key network components as well as different architectural options and characteristics of deployment scenarios. Acronyms: user equipment (UE), access point (AP), central unit (CU). Icons designed by Freepik.}
    \label{fig:CLOSE_6G}
\end{figure*}

Despite a large body of research on \ac{D-MIMO} communication and also on distributed radar, few studies on \ac{ISAC} \ac{D-MIMO} have been conducted. For instance, with proper optimization, \ac{ISAC} beamforming can reach similar performance as sensing-prioritized and communication-prioritized systems~\cite{MIMO-ISAC-arxiv}. It is also shown that one can deploy a cloud radio access network architecture to facilitate centralized \ac{ISAC} processing of all \acp{AP} \cite{behdad2022globecom}. In \cite{Loc_DMIMO_2022}, a downlink \ac{D-MIMO} system is studied from a positioning perspective. However, no previous study has provided a comprehensive vision of \ac{ISAC} in \ac{D-MIMO} systems and an analysis of the key challenges and opportunities of \ac{ISAC} in \ac{D-MIMO}.

In this paper, we investigate the potentials and challenges of \ac{D-MIMO} networks providing \ac{ISAC} operations, referred to as \ac{ISAC} \ac{D-MIMO}. As illustrated in Fig. \ref{fig:CLOSE_6G}, we assume that a set of cooperative multi-antenna \acp{AP} with different capabilities perform communication, localization, and sensing jointly, possibly within the same spectrum resources. We introduce the architecture requirements of \ac{ISAC} \ac{D-MIMO}  networks and present the key open problems to be addressed in such distributed multi-functional networks. Also, simulation results, as well as initial testbed evaluations, are presented. As demonstrated, the distributed and cooperative characteristics of \ac{D-MIMO} networks enable efficient joint communication, localization, and sensing, with reduced coordination requirements. We reveal that there are still multiple open problems to be addressed before such systems can be implemented in practice.   

\section{Deployments and Architectures of D-MIMO Networks}
\label{section:Architecture}
In this section, we present the deployment and architecture options for \ac{D-MIMO} networks. This also provides the basis for the architecture options of \ac{ISAC} \ac{D-MIMO} to be discussed in Section~\ref{section: Imple}.

Figure~\ref{fig:CLOSE_6G} illustrates \ac{ISAC} functionalities in \ac{D-MIMO} systems along the different architectural options and characteristics of deployment scenarios. A desirable \ac{D-MIMO} architecture is scalable, adaptive, and compatible with the current network standards allowing for seamless addition/removal of \acp{AP}\footnote{The terminology \ac{AP} in this paper may represent different levels of capability in the network implementation. For a more comprehensive terminology, please refer to~\cite{haliloglu2023dmimosystem}} with minimal network impact. Compatibility with legacy macro sites is also crucial, enabling efficient resource use and interference management while ensuring minimal impact on legacy \acp{UE}. A further exploration into each deployment and architectural option follows.

First of all, \ac{D-MIMO} is of interest in both indoor and outdoor deployments, with different use cases, objectives, and different kinds of connection options between the \acp{AP}. A possible use case is critical communications for indoor scenarios, e.g., in factories, warehouses, and offices, with support for dense machine-type communication or extended reality applications. In dense urban area scenarios, e.g., in airports, stadiums, public squares, outdoor \ac{D-MIMO} could still boost the capacity, where necessary, and provide coverage regardless of the site location and/or UE mobility. Additionally, a deployment with low visual footprints is desirable; radio stripes~\cite{Interdonato2019} are a candidate technology because they offer the possibility to hide the installation in existing infrastructure. 

Second, taking different deployment options into account, \ac{D-MIMO} is expected to support the spectrum ranging from sub-\qty{6}{GHz} to high bands. At low bands, e.g., \ac{FR1}, \ac{D-MIMO} can improve the \ac{SE} via, e.g., \ac{CJT}, which can also improve the sensing performance due to phase-coherent operation. At higher bands, e.g., \ac{FR2} and beyond, \ac{D-MIMO} can be used to improve the reliability/availability of the access links to the \acp{UE}, thanks to macro-diversity against blockers and the large available bandwidth resulting in high data rates even with low \ac{SE}. More specifically, the macro-diversity provided by \ac{D-MIMO} makes it possible to provide \ac{LOS} link(s) to the \ac{UE}s and reduce the effect of high penetration loss especially at high bands. These  \ac{LOS} links are required for many \ac{LAS} applications. 

Third, for both centralized and distributed processing, the fronthaul (between the \ac{CU} and \ac{AP}) and the backhaul (between \ac{CU} and network) requirements depend on the number of \acp{UE}, the deployment of \acp{CU}/\acp{AP}, their processing capabilities, and the supported operation modes of the \ac{D-MIMO} network. The goal is to reduce the required processing at the nodes close to the \acp{UE}, reducing their cost, complexity, and simplifying deployment, which
in turn increases the fronthaul/backhaul traffic. Also, it is interesting to note that the processing at higher frequency bands, desirable for communications, makes the hardware (e.g., converters) more power-hungry. As a result, a large part of the processing functionalities may be moved to the \acp{CU} instead of the \acp{AP}, at the cost of fronthaul/backhaul traffic. At low frequencies, on the other hand, processing can be performed at the \acp{AP}, resulting in low fronthaul/backhaul requirements. The operation modes, ranging from \ac{CJT} with centralized processing (having highest efficiency and fronthaul/backhaul requirement)  to non-\ac{CJT} with duplicated data in each \ac{AP} (having lowest efficiency and fronthaul/backhaul requirement), depend on the considered use cases. Within fronthaul/backhaul transport, the focus is on different lower-layer split options balancing \ac{AP} and fronthaul complexity, cost, and capacity, while achieving high performance. 

Fourth, the fronthaul/backhaul transport medium is expected to be based on a combination of fiber and wireless for both communication and \ac{LAS}. Fiber is preferred, when feasible, while wireless fronthaul/backhaul provides increased flexibility and short time-to-market. Here, the so-called \ac{IAB} \cite{9187867}, with the access and backhaul connections provided in the same node, is a candidate solution with flexible access/fronthaul/backhaul resource allocation capability. 

Fifth, full-duplex operations may improve communication performance, compared to half-duplex systems using, e.g., dynamic time division duplex. Full-duplex also enables mono-static sensing, similar to some conventional radars. To integrate sensing, localization, and communication, full-duplex operation, or very short \ac{DL}/\ac{UL} switching delays, could potentially be beneficial. However, in practice, self-interference cancellation is challenging, since, in general, more than \qty{100}{dB} isolation may be required, which imposes strict hardware capability requirements. However, proper deployment of antenna panels and/or beamforming can partly reduce the effect of self-interference, and dense deployment of \ac{D-MIMO} \acp{AP} will substantially lower the power difference of the transmitted/received signals.

The sixth and final architectural option relates to phase synchronization in \ac{D-MIMO}, which enables the alignment of signal phases across multiple distributed antennas, i.e., establishing \textit{phase coherence}. This ensures that signals combine constructively at the \acp{AP} and \acp{UE} to achieve the desired array gain.  Phase synchronization is essential for \ac{CJT} and it is easier to achieve at low frequencies. It is likely that, at least in the early roll-outs of \ac{D-MIMO}, non-coherent transmission will be considered at high frequencies for both communication and \ac{LAS}. On the other hand, at low frequencies, over-the-air calibration methods can be applied to enable phase-aligned reciprocity-based beamforming across \acp{AP}.

\section{A Multi-Functional View of D-MIMO}
\label{section: Imple}
Based on the \ac{D-MIMO} architectures and deployments described in Section~\ref{section:Architecture}, in this section, we discuss how communication, localization, and sensing tasks can be accomplished and how these services can benefit from D-MIMO.

\subsection{D-MIMO from a Communication Perspective}
Some of the opportunities and challenges of \ac{D-MIMO} for communications are as follows:
{\begin{itemize}[leftmargin=*]
    \item \emph{Indoor and outdoor considerations:} With the key idea of planning the network around the user rather than the cell, \ac{D-MIMO} is more desired for indoor communications with high-quality performance requirements from a high density of \acp{UE}. In contrast to deploying massive \ac{MIMO} \acp{AP} outdoors and boosting the performance with spatial multiplexing and beamforming, indoor \ac{D-MIMO} is applied to realize a logical ``array'' with \acp{AP} equipped with a small number of antennas to improve the capacity and throughput.
    \item \emph{Operational bands:} One opportunity for \ac{D-MIMO} is multi-band operations where, for instance, depending on the traffic model, service requirement, etc., some \acp{AP} may operate at low or high bands for different purposes. For instance, assume a highway scenario with a large number of vehicles at low speeds in the morning, and few vehicles at high speeds during the night. In this situation, the network experiences diverse quality-of-service requirements, sensing/communication priorities, etc., during the day. Here, the presence of multiple nodes gives flexibility for multi-band operation. 
    \item \emph{Centralized and distributed processing: } Distributed or centralized processing is based on the \acp{AP} capabilities. For instance, depending on the operational frequency, their associated processing can be in the \acp{CU} or \acp{AP}. Distributed antenna deployment provides broader resource trade-off options based on local data traffic and nodes' deployment. Moreover, the existence of multiple nodes opens opportunities for preventing service outages and reducing self-interference, through optimized deployment and coordinated beamforming techniques.
    \item \emph{Fronthaul and backhaul:} Wireless fronthaul/backhaul is preferred outdoors with low cost and fast deployment, whereas wired fronthaul/backhaul could be more beneficial indoors with improved reliability and capacity. Also, at higher operation bands, wireless deployment is preferred with less restricted synchronization requirements. Moreover, some benefits provided by the architecture of fronthaul/backhaul in \ac{D-MIMO} networks include: 
    \begin{itemize}
    \item Cooperative communications. Here, the presence of fronthaul/backhaul helps the nodes to have multiple views on the \ac{UE}/object which improves the \ac{CSI} quality significantly; see Fig.~\ref{fig:JSAC_DMIMO} as an example.  
    \item Multi-band operation. Here, the nodes can operate in different bands, obtain information, and then share them via fronthaul/backhaul. This is advantageous in terms of interference mitigation and resource allocation. 
    \item Scalability. Different protocol layers should support the scalability requirements where, similarly as in \ac{IAB} networks \cite{9187867}, as long as a node is properly connected to its ``parent'' node(s), it does not need to be connected (in terms of management) to the rest of the network, since the parent node implements the control functions. Also, while some long-term management may be handled by the \ac{CU}(s), a large part of the dynamic decisions of each \ac{UE} can be handled by its serving \ac{AP}(s). 
    \end{itemize}
    \item \emph{Half- and Full-duplex:} Theoretically, full-duplex is preferred for communication because it almost doubles the \ac{SE}; however, the existing problem with self-interference at the \ac{AP}, i.e., the interference between the transmit and receive antenna arrays, could reduce the expected \ac{SE} gains significantly. In \ac{D-MIMO} systems, full-duplex could provide more flexibility in terms of, e.g., channel estimation and interference coordination.
    \item \emph{Coherent and non-coherent processing:}  Coherent-phase synchronization refers to the process of aligning the phase of the signals transmitted or received by different \acp{AP}. This alignment is crucial in coherent \ac{D-MIMO} systems because it ensures that the signals from different \acp{AP} interfere constructively, maximizing the signal strength at the receiver. Time synchronization, on the other hand,  ensures that the signal arrivals from different antennas are synchronized in time, which is essential for decoding the signals correctly, especially in systems employing advanced signal processing techniques like beamforming. Non-coherent processing has a lower cost and might be sufficient for narrow beams and spatial multiflow, while coherent processing could improve the reliability at the cost of increased complexity.
\end{itemize}}

\subsection{{D-MIMO from a Localization and Sensing Perspective}}
In the \ac{D-MIMO} context, sensing can rely on \ac{DL} or \ac{UL} pilots. \ac{UL} pilots are more compatible with standard \ac{D-MIMO} processing, as they are used for channel estimation and reciprocity-based \ac{DL} precoding. On the other hand, orthogonal \ac{DL} pilots are preferred since they can allow the same pilots to be reused efficiently for all \acp{UE}. \Acp{AP} can receive and process \ac{DL} transmissions from other \acp{AP}, (providing opportunities for bi- and multi-static sensing) or from themselves (for mono-static sensing). In terms of localization, both \acp{UE} and \acp{AP} may need to be localized, where the latter type can be seen as a calibration (e.g., when a new \ac{AP} is added to an existing system). 

With this background, we can now consider the architectural dimensions.
{\begin{itemize}[leftmargin=*]
       \item \emph{Indoor and outdoor considerations:} Indoor scenarios are challenging as they have more clutter, which causes more multipath that affects localization accuracy and missed detection of the wanted target in sensing, while the high path-loss and high mobility may limit outdoor performance. To remove or suppress the interference of clutter, either more bandwidth or novel signal processing is needed, while novel waveforms and/or processing are needed to support high mobility. The indoor and outdoor deployments are likely to differ. For instance, from a localization perspective, it is desirable to have \acp{AP} distributed at different heights to estimate the elevation of the \ac{UE}, which could be seen as more important for outdoor use cases.
        \item \emph{Operational bands:} The low-frequency ranges, i.e., in \ac{FR}1, have a rich multipath profile, which makes it harder to perform \ac{LAS} due to multipath interference. On the other hand, the possibility of phase-coherent processing provides a means to resolve multipath and attain high accuracy. A promising alternative is the use of machine learning at lower frequencies in the form of fingerprinting. Higher frequency ranges have a more sparse multipath profile and larger available bandwidth, providing a direct way to reject multipath interference. However, at \ac{FR2} and above, phase synchronization may not be attainable, so we revert to classical \ac{LAS} methods. To some extent in \ac{FR1}, but especially in \ac{FR2}, \ac{LOS} blockage detection will play an important role, as each receiver may be associated with a large number of transmitters but only a subset of which will have a \ac{LOS} condition. 
        \item \emph{Centralized and distributed processing: } Three important scalability aspects should be considered, namely update rate, transmission, and processing. \Ac{LAS} are low-rate services, requiring periodic activation at a low rate of $10$ or $\qty{100}{Hz}$, depending on the application and mobility. This means that they allow flexible scaling with the number of users or objects to be tracked. In terms of transmission, the transmitters should apply orthogonal waveforms, which require coordination in time and frequency. Consequently, \ac{LAS} pilot transmissions scale with the number of transmitters (e.g., \ac{UL} localization scales with the number of \acp{UE}, and multi-static sensing scales with the number of transmitting \acp{AP}). In terms of processing, \ac{DL} localization can be performed decentralized at each \ac{UE}, while data fusion from each receiver is needed for sensing and \ac{UL} localization, causing processing delays. Under non-coherent processing, it is sufficient to perform fusion based on the locally processed information (e.g., delay estimation for localization or locally detected objects for sensing), with a data requirement that scales with the number of objects. Under phase coherent fusion, the fusion is based on the raw \ac{IQ} data, which may correspond to a large amount of data. For sensing, it will also be important to separate the moving objects from the static background clutter, which requires extended integration times for Doppler estimation.
        \item \emph{Fronthaul and backhaul:} Precise time synchronization and/or phase synchronization between the \acp{AP} place additional demands on the fronthaul or backhaul, as wired links to a master clock must be installed or continuous operation over an over-the-air synchronization protocol must be provided. Similarly, sensing also requires time or phase synchronization for improved performance.
        \item \emph{Half- and full-duplex:} Localization does not require full-duplex devices. However, sensing with full-duplex capability could help localization with reduced beam training overhead. In the case of mono-static sensing, each \ac{AP} should be able to operate in full-duplex mode or be equipped with a well-isolated transmitter and receiver that share a common clock. For multi-static sensing, full-duplex is not a requirement, but the transmission of orthogonal signals should be well-coordinated.
        \item \emph{Coherent and non-coherent processing:} For delay-based positioning, precise time synchronization (sub ns-level) between the \acp{AP} is needed to relate the delay measurements to/from different \acp{AP}. If such synchronization is not possible, round-trip-time protocols can be used for positioning, while for sensing, \ac{LOS} paths can provide a timing reference. For phase-coherent processing in both \ac{LAS} tasks, even precise phase synchronization between the \acp{AP} must be attained, meaning that the signal phase at one \ac{AP} can be related to the \ac{UE} location and the signal phase at another \ac{AP}. This means that the phases should not only be fixed but also be perfectly known. The reason is that in \ac{LAS}, phase measurements are exploited to extract geometric information (delay and position) as opposed to communications where channel estimation inherently compensates for fixed phase offsets and thus circumvents the need for knowing the exact phase values. Hence, the phase center of each \ac{AP} must be determined and phase offsets, e.g., due to cables, must be calibrated. 
\end{itemize}}

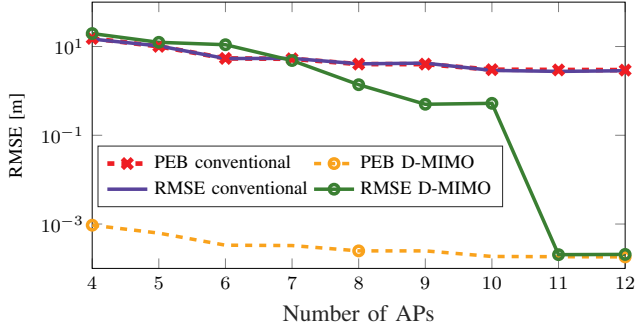
\begin{figure}[t]
        \centering
%
%
\definecolor{mycolor1}{rgb}{0.49400,0.18400,0.55600}%
\begin{tikzpicture}[scale=1\columnwidth/10cm,font=\footnotesize]
\begin{axis}[%
width=8cm,
height=4cm,
scale only axis,
xmin=4,
xmax=12,
xlabel style={font=\color{white!15!black}},
xlabel={Number of APs},
ylabel={RMSE [m]},
ymode=log,
ymin=0.0001,
ymax=1e2,
yminorticks=true,
 legend columns=2,
axis background/.style={fill=white},
legend style={at={(0.01,0.23)},
  anchor=south west,legend cell align=left, align=left, draw=white!15!black}
]
\addplot [color=Red, line width=2.0pt, dashed, mark=x, mark options={solid, Red}, mark size=3]
  table[row sep=crcr]{%
4	15.2176169919147\\
5	10.0913914550671\\
6	5.39339785317973\\
7	5.29125030909978\\
8	4.01452982482993\\
9	4.01447298871339\\
10	3.00559365953909\\
11	2.96598102096157\\
12	2.93889990256433\\
};
\addlegendentry{PEB conventional}

\addplot [color=YellowOrange, line width=1.5pt, dashed,mark=o,mark options={solid, YellowOrange},mark repeat=4]
  table[row sep=crcr]{%
4	0.000941345919631802\\
5	0.000624242953064326\\
6	0.000333629967473828\\
7	0.000327311226906815\\
8	0.000248334628992942\\
9	0.000248331113170036\\
10	0.000185922889831015\\
11	0.000183472493312913\\
12	0.000181797283566482\\
};
\addlegendentry{PEB D-MIMO}

\addplot [color=Violet, line width=1.5pt, mark options={solid, Violet}]
  table[row sep=crcr]{%
4	14.7016482069736\\
5	10.3186522144862\\
6	5.36080463974367\\
7	5.40317575305279\\
8	4.07125856657428\\
9	4.2204510300727\\
10	2.8972590001391\\
11	2.74978367692641\\
12	2.84713512865197\\
};
\addlegendentry{RMSE conventional}

\addplot [color=OliveGreen, line width=1.5pt, mark=o, mark options={solid, OliveGreen}]
  table[row sep=crcr]{%
4	19.6007869596161\\
5	12.4333161572287\\
6	10.9765013138861\\
7	4.83543537827475\\
8	1.37900218373112\\
9	0.499791604362087\\
10	0.524611712286341\\
11	2.0420e-04\\
12	2.0673e-04\\
};
\addlegendentry{RMSE D-MIMO}

\end{axis}
\end{tikzpicture}%
        \vspace{-0.8cm}
        \caption{Impact of the number of \acp{AP} on \ac{PEB} and positioning \ac{RMSE}, comparing conventional time-coherent positioning with phase-coherent \ac{D-MIMO} positioning. The system operates at 28 GHz with 6 MHz of bandwidth, under pure \ac{LOS} conditions. }
        \vspace{-6mm}
        \label{fig:numberofBS}
        \end{figure}

As an example, Fig.~\ref{fig:numberofBS} shows the impact of the number of \acp{AP} on \ac{PEB} and positioning \ac{RMSE}, comparing conventional time-coherent positioning with phase-coherent \ac{D-MIMO} positioning. This shows the theoretical benefits of a D-MIMO solution for accurate positioning. In terms of \ac{PEB}, \ac{D-MIMO} performance outperforms the corresponding conventional \ac{PEB} by several orders of magnitude. In terms of \ac{RMSE}, the gap would disappear when a sufficient number of connected \acp{AP} are present. Otherwise, the ambiguities due to the use of carrier phase limit the performance.

 \newcommand{\hwplotgreen}{\raisebox{-1pt}{\tikz{\draw (0,0) node[minimum height=0.2cm, minimum width=0.5cm, draw,fill=green!5!white] {};}}}
\newcommand{\hwplotblue}{\raisebox{-1pt}{\tikz{\draw (0,0) node[minimum height=0.2cm, minimum width=0.5cm, draw,fill=blue!5!white] {};}}}

\begin{table*}
         
    \caption{Suitability Matrix of Architectural Options Combinations with Implementation Comments: Evaluating Communication and Localization \& Sensing. 
} 
\vspace{-0.2cm}
    \centering
{\hwplotgreen} : Both options are feasible/possible \quad  {\hwplotblue} : One of the options is preferable 

\vspace{0.2cm}
    \begin{tabular}{|Q{1.28cm}|Q{7.66cm}|Q{7.66cm}|}
    \hline
    \rowcolor{blue!0}
     \textbf{} & \textbf{Indoor vs. Outdoor } & \textbf{Higher bands vs. Lower bands} \\
    \hline
\cellcolor{blue!0} 
\makecell{\textbf{Centralized} }
& \cellcolor{green!5} \makecell[lb]{\textbf{Both options} are \textit{feasible} \\
\textbf{Com.}: improves spectral efficiency/dynamic blocking mitigation\\
\textbf{L\&S}: suitable for  time- and phase-coherent processing, \\only lower mobility
}
 & \cellcolor{blue!5} \makecell[lb]{\textbf{Higher bands} \textit{preferred} (dense APs and low cost)\\
\textbf{Com.}: fast control of narrow beams, \\but high requirements on backhaul/fronthaul\\
\textbf{L\&S}: phase-coherent capability
}
\\
\hline
\cellcolor{blue!0}\textbf{Distributed} 
& \cellcolor{green!5} \makecell[lb]{\textbf{Both options} are \textit{possible} \\
\textbf{Com.}: improves scalability (less reliability for indoor)\\
\textbf{L\&S}: suitable for time-coherent processing
}
 & \cellcolor{blue!5} \makecell[lt]{\textbf{Lower bands} \textit{preferred} (less dense APs/high resolution converters)\\
\textbf{Com.}: Lower data rates allow for more advanced APs,\\ resulting in low backhaul requirements, \\but interference might limit spectral efficiency\\
\textbf{L\&S}: suitable for time-coherent processing
}
\\
\hline
\cellcolor{blue!0}\textbf{Wireless front- and backhaul} 
& \cellcolor{blue!5} \makecell[lb]{\textbf{Outdoor} \textit{preferred} (\textit{less blockers})\\
\textbf{Com.}: low cost and fast deployment, but less reliability \\
\textbf{L\&S}: time-coherent processing
}
 & \cellcolor{blue!5} \makecell[lb]{\textbf{Higher bands} \textit{preferred}, (\ac{IAB} feasible, \\less restricted sync requirements)
 \\
\textbf{Com.}: low cost and fast deployment, but less reliability\\ 
\textbf{L\&S}: time-coherent processing
}
\\
\hline
\cellcolor{blue!0}\textbf{Wired front- and backhaul} 
& \cellcolor{blue!5} \makecell[lb]{\textbf{Indoor} \textit{preferred} (might be costly for outdoors)
\\
\textbf{Com.}: improves reliability and backhaul fronthaul capacity\\
\textbf{L\&S}: supports tight sync requirements for \\phase-coherent processing
}
 & \cellcolor{green!5} \makecell[lb]{\textbf{Both options} are \textit{feasible} \\ 
\textbf{Com.}: improves reliability and backhaul fronthaul capacity,\\ important especially in higher bands\\
\textbf{L\&S}: not needed for higher bands, except for certain challenging  \\cases or use of \ac{AI}
}
\\
\hline
\cellcolor{blue!0}\textbf{Half-Duplex} 
& \cellcolor{green!5} \makecell[lb]{\textbf{Both options} are \textit{possible}\\
\textbf{Com.}: lower cost, but increased delays\\
\textbf{L\&S}: suitable for  time- and phase-coherent processing
}
 & \cellcolor{green!5} \makecell[lb]{\textbf{Both options} are \textit{possible} \\
\textbf{Com.}: lower cost, but increased delays \\
\textbf{L\&S}: suitable for time-coherent processing (higher bands)\\phase-coherent processing (lower bands)
}
 \\
\hline
\cellcolor{blue!0}\textbf{Full-Duplex} 
& \cellcolor{blue!5} \makecell[l]{\textbf{Indoor} \textit{preferred} (due to low transceiver signal strength difference)\\
\textbf{Com.}: lower latency\\
\textbf{L\&S}: enables monostatic sensing, severe leakage challenges
}
 &  \cellcolor{blue!5} \makecell[lb]{\textbf{Higher bands} \textit{preferred}  (due to beam-based spatial\\ transceiver isolation and short hops)\\
\textbf{Com.}: flexible \ac{TDD} deployment\\
\textbf{L\&S}: enables monostatic sensing, severe leakage challenges
}
 \\
\hline
\cellcolor{blue!0}\textbf{Non-coherent} 
& \cellcolor{green!5} \makecell[lb]{\textbf{Both options} are \textit{possible}\\
\textbf{Com.}: lower cost but might result in insufficient reliability\\
\textbf{L\&S}: suitable for time-coherent processing
}
 & \cellcolor{blue!5} \makecell[lb]{\textbf{Higher bands} \textit{preferred} (low spectral efficiency and resolution\\ in Lower bands)\\
\textbf{Com.}: lower cost and might be sufficient \\for narrow beams and spatial multiflow \\
\textbf{L\&S}: suitable for time-coherent processing
}
 \\
\hline
\cellcolor{blue!0}\textbf{Coherent} 
& \cellcolor{blue!5} \makecell[lb]{\textbf{Indoor} \textit{preferred} (due to short inter-\ac{AP} distances)\\
\textbf{Com.}: improves reliability\\
\textbf{L\&S}: suitable for  time- and phase-coherent processing
}
 & \cellcolor{blue!5} \makecell[lb]{\textbf{Lower bands} \textit{preferred} (due to lower carrier frequency)\\
\textbf{Com.}: improves reliability\\
\textbf{L\&S}: suitable for  time- and phase-coherent processing}
\\
\hline
    
    \end{tabular}
\label{tab:II}
\end{table*}

\section{Toward ISAC in D-MIMO: Potentials and Implementations}
\label{section: challenge}
In this section, we consider a converged \ac{ISAC} \ac{D-MIMO} system, from four perspectives:
\begin{inparaenum}[i)]
    \item the architecture and deployment;
    \item standardization;
    \item quantitative benefits of \ac{ISAC} \ac{D-MIMO}; and 
    \item implementation challenges.
\end{inparaenum}
\subsection{ISAC D-MIMO Architectures}
Based on the discussions from Sections~\ref{section:Architecture} and~\ref{section: Imple}, and with specific focus on the indoor vs. outdoor and higher vs. lower bandwidth options, we summarize in Table~\ref{tab:II} the main implications of the different architectural options. Green blocks indicate that both scenarios are possible/feasible, while blue blocks show that there is a preferred deployment. Each block assesses the suitability for communication and \ac{LAS}. It is evident that for \ac{D-MIMO} communication, the favored architecture encompasses phase-coherent distributed processing, half-duplex, and wired fronthaul and backhaul, particularly for outdoor environments in the \ac{FR}1. For \ac{LAS}, while half-duplex and wired connections are favored, interest also extends to both distributed non-coherent \ac{FR}2 and centralized coherent \ac{FR}1 operations across indoor and outdoor settings. Hence, a preferred \ac{ISAC} \ac{D-MIMO} architecture mirrors the preferred communication architecture but incorporates centralized processing, such as phase-coherent \ac{IQ} samples sharing for  \ac{LAS}. 

\subsection{ISAC D-MIMO Standardization}
Communication networks primarily rely on standardized operations, while sensing signal processing methods are based on proprietary, i.e., non-standardized solutions. On the other hand, with \ac{ISAC}, the transmitted signals for ISAC, supporting both communication and sensing functions (either simultaneously or time/frequency/space multiplexed) require standardization, as well as the associated control signaling. In some sense, these considerations are general for all 6G ISAC technologies. What sets \ac{D-MIMO} apart is the multi-static sensing perspective, considering several concurrent \ac{AP} transmitters and/or several concurrent \ac{AP} receivers. Again, processing will be proprietary, but signal design and coordination will rely on standardized solutions. This necessitates extensive standardization efforts to incorporate sensing into D-MIMO. For instance, the current 3GPP standardization on multi-\acp{AP} concentrates mainly on the case of ideal backhaul/synchronization, but work on enhancements for non-ideal operation has started in 3GPP Rel-19. 3GPP started preliminary discussions on \ac{ISAC} from Rel-19 in early 2024.

\subsection{ISAC D-MIMO Quantitative Benefits---A Case Study}
Communication, localization, and sensing can operate harmoniously in ISAC D-MIMO. As an example, we consider a scalable \ac{D-MIMO} simulation scenario, assuming perfect time and phase synchronization between \acp{UE} and \acp{AP}. Fig.~\ref{fig:JSAC_DMIMO} shows the sum \ac{UL} \acp{SE} as a function of transmit \ac{SNR} averaged over different \ac{UE} locations and shadow fading realizations. {With the setups shown in the caption, \ac{MRC} is used to leverage channel estimations in various scenarios where sensing is used to detect blockage status, while localization is used for \ac{CSI} estimation (assuming a prior radio map exists):  
    \emph{(i) with ISAC:} Having both blockage status information and \ac{CSI}, the \acp{UE} are assigned to \acp{AP} without \ac{AP}-\ac{UE} blockage with perfect \ac{CSI}; 
    \emph{(ii) with localization:} The \acp{UE} have perfect \ac{CSI} but without the information of blockage from sensing, they are still served by the default \acp{AP}; 
    \emph{(iii) with sensing:} The \acp{UE} are assigned to the back-up \acp{AP} but with no \ac{CSI} from localization; 
    \emph{(iv) without ISAC:} The \acp{UE} are served by default \acp{AP} without \ac{CSI}.
As shown in Fig. \ref{fig:JSAC_DMIMO}, \ac{LAS} improves the \ac{UL} \ac{SE} considerably. For instance, consider the parameter settings of Fig. \ref{fig:JSAC_DMIMO} and \ac{SNR} 10 dB. Then, compared to the cases without \ac{ISAC}, adding localization (i.e., providing \ac{CSI}), sensing (i.e., providing knowledge about blockage), and both \ac{LAS}, improve the UL SE by $8 \times$, $12\times$, and $16 \times$, respectively.

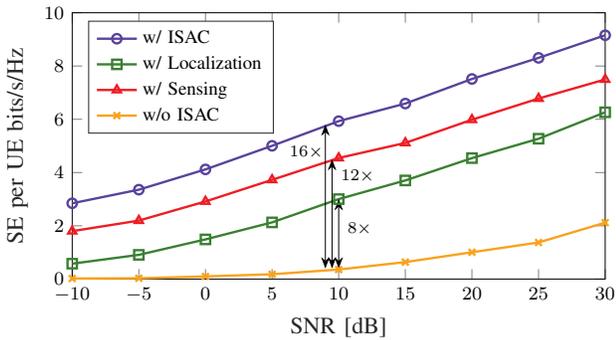
\begin{figure}
    \centering
%
%
\definecolor{mycolor1}{rgb}{0.00000,0.44700,0.74100}%
\definecolor{mycolor2}{rgb}{0.85000,0.32500,0.09800}%
\definecolor{mycolor3}{rgb}{0.92900,0.69400,0.12500}%
\definecolor{mycolor4}{rgb}{0.49400,0.18400,0.55600}%
\begin{tikzpicture}[scale=1\columnwidth/10cm,font=\footnotesize]
\begin{axis}[%
width=8cm,
height=4cm,
scale only axis,
scale only axis,
xmin=-10,
xmax=30,
xlabel style={font=\color{white!15!black}},
xlabel={SNR [dB]},
ymin=0,
ymax=10,
ylabel style={font=\color{white!15!black}},
ylabel={SE per UE bits/s/Hz},
axis background/.style={fill=white},
legend style={at={(0.03,0.97)}, anchor=north west, legend cell align=left, align=left, draw=white!15!black}
]
\addplot [color=Violet, line width=1.0pt, mark=o, mark options={solid, Violet}]
  table[row sep=crcr]{%
-10	2.84502420984444\\
-5	3.35641754437544\\
0	4.11725777204359\\
5	5.0037556092651\\
10	5.92957697025488\\
15	6.58994339375756\\
20	7.51372677977619\\
25	8.3064001399664\\
30	9.15581879275652\\
};
\addlegendentry{w/ ISAC}

\addplot [color=OliveGreen, line width=1.0pt, mark=square, mark options={solid, OliveGreen}]
  table[row sep=crcr]{%
-10	0.577268341826918\\
-5	0.908486386560689\\
0	1.48832748438636\\
5	2.12770492708098\\
10	3.00220280217525\\
15	3.70303057875158\\
20	4.54181096964026\\
25	5.2695988864442\\
30	6.26240328681373\\
};
\addlegendentry{w/ Localization}

\addplot [color=Red, line width=1.0pt, mark=triangle, mark options={solid, Red}]
  table[row sep=crcr]{%
-10	1.79854618162879\\
-5	2.19537711044062\\
0	2.91393077199915\\
5	3.72573518433415\\
10	4.54572620939326\\
15	5.11221131746925\\
20	5.98355875143768\\
25	6.7793007200692\\
30	7.49506883122273\\
};
\addlegendentry{w/ Sensing}

\addplot [color=YellowOrange, line width=1.0pt, mark=x, mark options={solid, YellowOrange}]
  table[row sep=crcr]{%
-10	0.0203884441840984\\
-5	0.0286574882783473\\
0	0.0985629541261956\\
5	0.179330135269899\\
10	0.357062008526195\\
15	0.635551248899955\\
20	1.00610055563391\\
25	1.37276946741008\\
30	2.1182784487073\\
};
\addlegendentry{w/o ISAC}

\draw[Stealth-Stealth] (axis cs: 10,0.4)   -. (axis cs: 10,3) node[midway, above right=-0.2cm and -0.2cm] {\scriptsize \begin{tabular}{c}
    $8 \times$ 
\end{tabular}};

\draw[Stealth-Stealth] (axis cs: 9.5, 0.4)   -. (axis cs: 9.5, 4.5) node[midway, above right= 0.3cm and -0.2cm] {\scriptsize \begin{tabular}{c}
    $12 \times$
\end{tabular}};

\draw[Stealth-Stealth] (axis cs: 9, 0.4)   -. (axis cs: 9,5.8) node[midway, above left= 0.4cm and -0.3cm] {\scriptsize \begin{tabular}{c}
    $16 \times $
\end{tabular}};

\end{axis}
\end{tikzpicture}%
    \caption{Impact of \ac{LAS} on the UL SEs in a simulated phase-coherent \ac{D-MIMO} system. We consider the setup from~\cite{cell-free2} {with $5$ UEs served by nearby \acp{AP} (in total $200$)} according to the dynamic cooperation clustering framework. The \acp{AP} and \acp{UE} are uniformly distributed in a $1\times 1$ km square area and the \acp{UE} are initially served by several \acp{AP} (default \acp{AP}) where the links are blocked. Rician fading channel model is used with the same parameter setup as in~\cite{ozdogan2019massive}.} 
    \vspace{-0.55cm}
    \label{fig:JSAC_DMIMO}
\end{figure}

\subsection{ISAC D-MIMO---Implementation Opportunities and Challenges}

\begin{figure*}
    \centering
 \includegraphics[width=1.99\columnwidth]{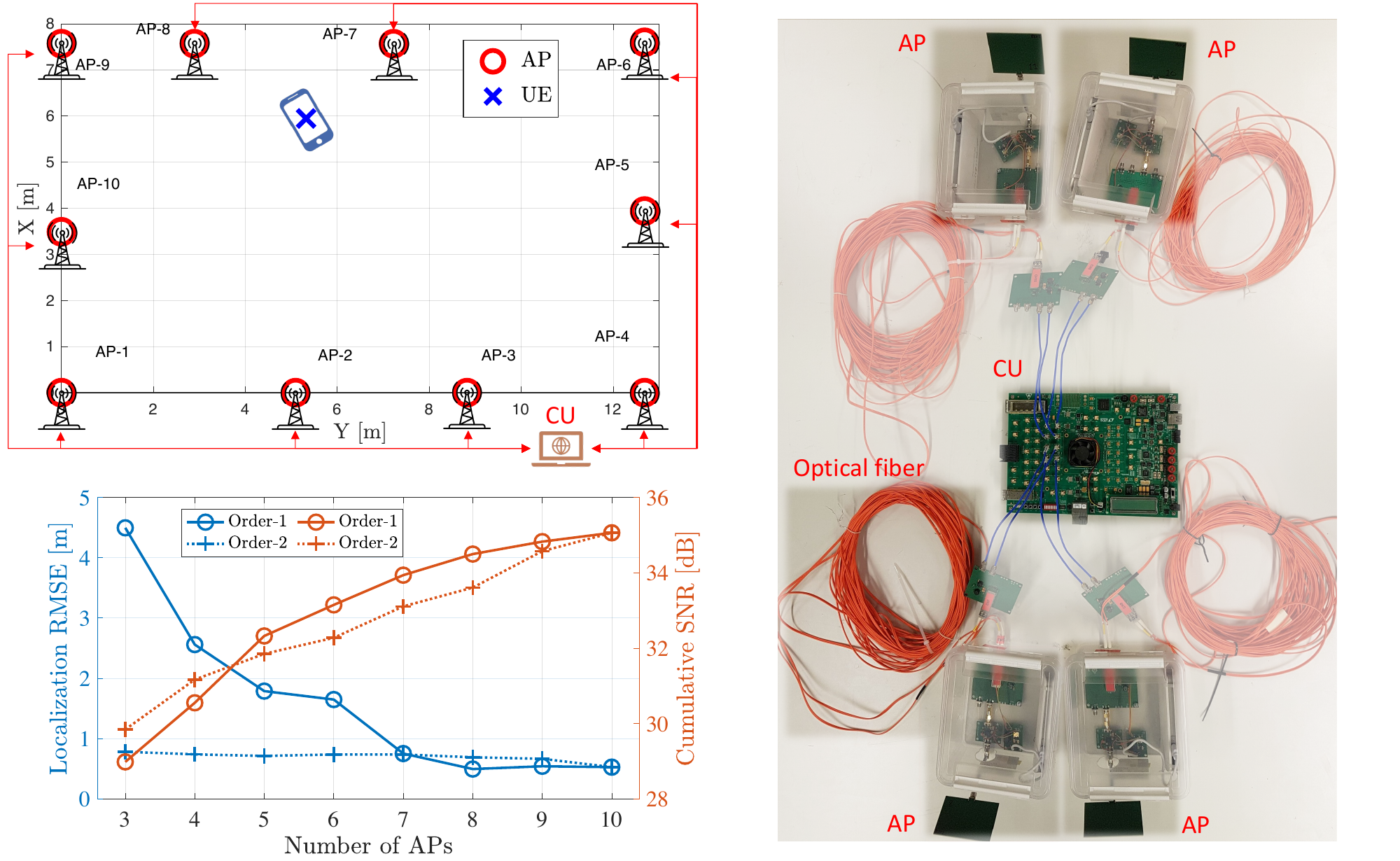}
    \caption{A \ac{D-MIMO} testbed used for ISAC demonstrations (right), the geometric configuration of the APs and the UE (upper left), and the experimental results for localization and communications (lower left). Localization RMSE and communication SNR performances are shown for different orders of deployment of APs (Order-1: $1~2~3~ 
 10~ 9~ 8~ 7~ 6~ 5~ 4 $, Order-2: $1~7~4~10~5~2~8~3~9~6$).}
    \label{fig:testbed_combined}
\end{figure*}

So far, only limited testbed activities exist involving \ac{D-MIMO} in general \cite{loschenbrand2022towards}, \ac{ISAC} in general~\cite{testbedcommag2023}, and \ac{ISAC} in \ac{D-MIMO}, in particular, \cite{testbedMar2024,callebaut22-02a,sezgin21-02a}. There is currently an urgent need to validate \ac{D-MIMO}, especially in conjunction with \ac{ISAC}. Two of the main technical challenges are \emph{scalability} and \emph{synchronization}~\cite{callebaut22-02a}. Scalability issues may arise because of fronthaul limitations, which may put a constraint on deployment configurations. Time and frequency synchronization is typically difficult to achieve, because of the distributed nature of the \acp{AP}, as elaborated on in Section~\ref{section:Architecture} and Section~\ref{section: Imple}. In \ac{D-MIMO} demonstrators, synchronization is typically achieved over the Ethernet or via dedicated cables; both solutions, however, result typically in non-scalable architectures~\cite{callebaut22-02a}.  A natural alternative is to perform synchronization over the air. However, this may result in significant overhead for certain deployment scenarios. A completely different approach for solving the synchronization problem is put forward in our testbed described in~\cite{sezgin21-02a} (see Fig.~\ref{fig:testbed_combined}), where phase synchronization issues are avoided by letting the \acp{AP} transfer to the \ac{CU} a $1$-bit quantized version of the analog radio-frequency signal via an optical cable. The advantage of this approach, which we refer to as 1-bit radio-over-fiber fronthaul, is that no local oscillators (which need to be synchronized for coherent transmission and reception) are present at the \acp{AP}. Furthermore, such a \ac{D-MIMO} architecture involves low cost \acp{AP} that can be built out of off-the-shelf components.  The disadvantage of this architecture is its limited scalability. 

Fig.~\ref{fig:testbed_combined} demonstrates a setup and the results of \ac{ISAC} experiments with the \ac{D-MIMO} 1-bit radio-over-fiber testbed \cite{Loc_DMIMO_2022}. The goal is to localize the \ac{UE} in \ac{DL} using known pilot signals from the fully synchronized \acp{AP}. We investigate the impact of \ac{AP} deployments on the performance of localization and communication, quantified by \ac{RMSE} and \ac{SNR}, respectively. Fig.~\ref{fig:testbed_combined} shows the \acp{RMSE} and cumulative \acp{SNR} as the number of \acp{AP} increases sequentially, considering two different orders for adding \acp{AP} to the \ac{D-MIMO} network, as stated in the figure caption. As expected, the geometric arrangement of the \acp{AP} (and the resulting \ac{GDOP}) plays a key role in localization accuracy, while it has a negligible impact on communication performance. Specifically, decreasing the number of \acp{AP} increases the sensitivity to the \ac{AP} locations for localization purposes, whereas its effect on location sensitivity in communication remains minimal. Thus, network planning can be simplified and flexible deployment can reduce the costs in \ac{D-MIMO} networks.

\section{Discussions and Outlook}
\label{section: Conclu}
D-MIMO and ISAC are set to be among the key enablers for 6G. This paper analyzed how the integration of ISAC in D-MIMO affects the underlying architecture. This analysis revealed both synergies and conflicts, while pointing towards D-MIMO architectures that can support all ISAC functionalities. We highlight preferred embodiments for communication and \ac{LAS}. Specifically, for communication, 
\begin{inparaenum}[i)]
    \item indoor, lower bands: coherent, wired backhaul, distributed processing, half duplex, and
    \item higher bands: noncoherent, centralized processing, half duplex (no wired/wireless backhaul preference, no indoor/outdoor preference) are preferred.
\end{inparaenum}
For \ac{LAS}, 
\begin{inparaenum}[i)]
   \item lower bands: coherent, wired backhaul, centralized processing, full duplex (no indoor/outdoor preference), and
   \item higher bands: noncoherent, distributed processing, full duplex (no wired/wireless backhaul preference, no indoor/outdoor preference) are desired. 
\end{inparaenum}

The paper also delved deeper into the quantitative performance benefits of ISAC in D-MIMO, from both \ac{LAS} and communication perspectives. These studies reveal significant synergies between communication and \ac{LAS}. Finally, the practical challenges of ISAC in D-MIMO implementation were considered, in particular, related to synchronization and scalability, highlighting the need for continued development in this area.

Overall, ISAC in D-MIMO seems promising to enhance system performance. However, there are still several open questions in D-MIMO that become further enriched by ISAC, especially related to scalability and adaptation in dynamic environments and efficient support of fast-moving users. Thus, ISAC in D-MIMO leads to a rich variety of D-MIMO and ISAC research problems, from both theoretical and practical perspectives.

\counting{
\section*{Acknowledgments}
This work was supported, in part, by the Gigahertz-ChaseOn Bridge Center at Chalmers in a project financed by Chalmers, Ericsson, and Qamcom. 
This work was also supported by the Swedish Foundation for Strategic Research (SSF), grants no. ID19-0021, ID19-0036, and  FUS21-0004,  the Swedish Research Council (VR grant 2022-03007, 2023-00272), the Chalmers Area of Advance Transport 6G-Cities project,  and the MSCA-IF grant 101065422 (6G-ISLAC). We would like to thank Dr. Yigeng Zhang for his help with the illustrations. }
}

\bibliographystyle{IEEEtran}
\bibliography{IEEEabrv,reference}

\begin{IEEEbiographynophoto}
{Hao Guo} is a Postdoc with the Department of Electrical Engineering at Chalmers University of Technology, Sweden.
\end{IEEEbiographynophoto}

\begin{IEEEbiographynophoto}
{Henk Wymeersch} is a Professor with the Department of Electrical Engineering at Chalmers University of Technology, Sweden.
\end{IEEEbiographynophoto}

\begin{IEEEbiographynophoto}
{Behrooz Makki} is a Senior Researcher at Ericsson, Sweden.
\end{IEEEbiographynophoto}

\begin{IEEEbiographynophoto}
{Hui Chen} is a Research Specialist with the Department of Electrical Engineering at Chalmers University of Technology, Sweden.
\end{IEEEbiographynophoto}

\begin{IEEEbiographynophoto}
{Yibo Wu} is a PhD Candidate with the Department of Electrical Engineering at Chalmers University of Technology, and a Researcher at Ericsson, Sweden.
\end{IEEEbiographynophoto}

\begin{IEEEbiographynophoto}
{Giuseppe Durisi} is a Professor with the Department of Electrical Engineering at Chalmers University of Technology, Sweden.
\end{IEEEbiographynophoto}

\begin{IEEEbiographynophoto}
{Musa Furkan Keskin} is a Research Specialist with the Department of Electrical Engineering at Chalmers University of Technology, Sweden.
\end{IEEEbiographynophoto}

\begin{IEEEbiographynophoto}
{Mohammad H. Moghaddam} is a R\&D Specialist at Qamcom Research and Technology, Gothenburg, Sweden. 
\end{IEEEbiographynophoto}

\begin{IEEEbiographynophoto}
{Charitha Madapatha} is a PhD Student with the Department of Electrical Engineering at Chalmers University of Technology, Sweden.
\end{IEEEbiographynophoto}

\begin{IEEEbiographynophoto}
{Han Yu} is a Postdoc with the Department of Electrical Engineering at Chalmers University of Technology, Sweden.
\end{IEEEbiographynophoto}

\begin{IEEEbiographynophoto}
{Peter Hammarberg} is a Senior Researcher at Ericsson, Sweden.
\end{IEEEbiographynophoto}

\begin{IEEEbiographynophoto}
{Hyowon Kim} is an Assistant Professor with the Department of Electronics Engineering at Chungnam National University, Daejeon, South Korea.
\end{IEEEbiographynophoto}

\begin{IEEEbiographynophoto}
{Tommy Svensson} is a Professor with the Department of Electrical Engineering at Chalmers University of Technology, Sweden.
\end{IEEEbiographynophoto}
\end{document}